\documentclass[aps,pra,twocolumn,superscriptaddress,nofootinbib]{revtex4-2}

\usepackage{graphicx}
\usepackage{amsmath,amssymb}
\usepackage{booktabs}
\usepackage{hyperref}
\hypersetup{colorlinks=true,linkcolor=blue,citecolor=blue,urlcolor=blue}

\usepackage{qcircuit}
\usepackage{tikz}
\usetikzlibrary{arrows.meta}

\newcommand{\Qx}{Q_x}
\newcommand{\Qz}{Q_z}
\newcommand{\rGHZ}{r_{\mathrm{GHZ}}}
\newcommand{\rpair}{r_{\mathrm{pairwise}}}
\newcommand{\Favg}{F_{\mathrm{avg}}}
\newcommand{\ket}[1]{\lvert #1\rangle}

\begin{document}

\title{Pauli Error Composition Determines the Multipartite Advantage in Conference Key Agreement}

\author{Ryder Carlson}
\email{rydercarlson26@gmail.com}
\affiliation{The Coding School dba Frontier Technology Institute, Los Angeles, California, USA}
\author{Lochlan Fitzgerald}
\affiliation{The Coding School dba Frontier Technology Institute, Los Angeles, California, USA}
\author{Kidus Peter}
\affiliation{The Coding School dba Frontier Technology Institute, Los Angeles, California, USA}
\author{Wilson Smith}
\thanks{mentor: helped formulate the question, supervised research, advised on manuscript organization}
\affiliation{Department of Computer Science, University of Maryland, College Park, Maryland, USA}

\date{\today}

\begin{abstract}
GHZ-based conference key agreement distributes a shared key to $N$ parties in one network use, whereas pairwise BB84 needs $N-1$. The survival of that advantage has been studied against party number, loss, and topology, but almost always with the channel fixed at depolarizing. Simulating both protocols exactly for $N = 3, 4, 5$ under dephasing, amplitude damping, and depolarizing noise, we find that the crossover is strongly dependent on the channel's bit- and phase-error contributions: normalized to average infidelity, amplitude damping and depolarizing agree to within 1.2\% at each group size, while dephasing can tolerate roughly $2.4\times$ more at $N = 3$.
We derive closed-form rate expressions for independent per-party Pauli noise and establish a penalty for complete noise concentration relative to an even split under fixed aggregate budgets in the low-noise regime. We also identify an optimal dealer assignment, yielding an approximately 18\% rate improvement in an asymmetric three-party example. One structural fact drives these results: the phase-error rate depends on a product over all $N$ arms, while each pairwise bit-error rate depends on only two arms.
\end{abstract}

\maketitle

\section{Introduction}\label{sec:intro}

Quantum key distribution creates a shared secret key between parties that is secured by physical law instead of computational assumptions. The two-party case is well developed, both theoretically and in deployed systems. When more than two parties require a common key, or a conference key, the protocol can be generalized in two ways. The parties can run bipartite QKD pairwise and combine the results, or they can distribute an $N$-party GHZ state and measure it directly, as in the $N$-party BB84 protocol of Grasselli et al.~\cite{grasselli2018}. The multipartite route has a resource argument in its favor, because in a bottleneck network where all parties connect through a central station, one GHZ state serves every party in a single network use, while the pairwise scheme requires $N-1$ separate distributions. Epping et al.\ formalized this advantage \cite{epping2017}, and it has been demonstrated experimentally over metropolitan distances \cite{proietti2021}.

The advantage is not unconditional, as GHZ states are more fragile than Bell pairs. The literature has mapped several ways the advantage can disappear: Memmen et al.\ show that transmission loss eliminates it beyond roughly 15~km per link, and threshold error rates fall as party number grows \cite{memmen2023}. Nearly all of these analyses, however, fix the channel at depolarizing, a uniformity with structural roots in how security proofs symmetrize noise (Section~\ref{sec:related}). As a result, the dependence of the crossover on channel type, as opposed to noise magnitude, party number, loss, or topology, has not been mapped.

This distinction is important because the two entropy terms in the conference key rate have different structure. The GHZ rate pays $h(\Qx)$, built from $\langle X^{\otimes N} \rangle$ and therefore a product over all $N$ arms, and $\max_i h(Q_{1,i})$, built from $\langle Z_1 Z_i \rangle$ and confined to a single pair. A channel that produces only phase-type errors leaves the second term identically zero, but one that produces bit-flip-type errors as well activates it. This suggests that the crossover depends on which Pauli components a channel produces, and that a threshold measured for depolarizing noise will transfer to some channels and badly miss others.

We test this by simulating both protocols exactly for $N = 3, 4, 5$ under dephasing, amplitude-damping, and depolarizing channels, locating the crossover by bisection, and normalizing to average channel infidelity so that thresholds are comparable across parameterizations. Four results come out of this. The raw thresholds spread by a factor of 8, which initially looks like a large channel effect; normalized to average infidelity, most of that spread is parameterization, and what remains sorts by error composition: amplitude damping and depolarizing, despite distinct Kraus structures and with both bit- and phase-type error contributions, cross within 1.2\% of each other at every group size, while dephasing tolerates $2.38\times$ more infidelity at $N = 3$.
Admixing bit-flip-type error into dephasing collapses the threshold by a factor of 14, front-loaded because it reflects the activation of a term rather than the growth of one. Thresholds fall with $N$, confirming earlier depolarizing-specific results under a per-qubit noise model. And at fixed total noise, concentration on one arm is significantly worse than an even split, which we support analytically using closed-form expressions for independent Pauli noise at fixed aggregate error budgets; this answers a question Proietti et al.\ posed numerically \cite{proietti2021} and yields a zero-cost operational rule for which party should hold the dealer role.

Section~\ref{sec:background} fixes notation and derives the structural asymmetry the results turn on. Section~\ref{sec:related} situates the work in the existing comparison literature. Section~\ref{sec:methods} describes the simulation. Section~\ref{sec:results} reports results, and Section~\ref{sec:discussion} discusses their common mechanism, the relation to prior work, and the limitations, chiefly the omission of transmission loss.

\section{Background}\label{sec:background}

\subsection{Notation}\label{sec:notation}

Throughout this paper, $N$ refers to the number of parties. Party 1 is the dealer (Alice), and parties 2 through $N$ are receivers (Bobs). We write the binary entropy function as
\begin{equation}
  h(x) = -x \log_2 x - (1-x)\log_2(1-x),
  \label{eq:binent}
\end{equation}
with $h(0) = h(1) = 0$. A single channel-strength parameter $\gamma \in [0, 1]$ controls the noise on each arm. The physical meaning of $\gamma$ depends on the channel and is given in Section~\ref{sec:channels}.

\subsection{Bipartite QKD and the entanglement-based baseline}\label{sec:bipartite}

In BB84 \cite{bennett1984}, Alice sends qubits prepared randomly in the $Z$ or $X$ basis, and Bob measures in a random basis. After sifting, the parties estimate the quantum bit error rate in each basis and extract a secret key at the asymptotic rate $r_{\mathrm{BB84}} = 1 - h(\Qz) - h(\Qx)$. The two entropy terms are the two costs of turning a noisy raw string into a secret one. The first, $h(\Qz)$, is the information leaked during error correction, since the parties must publicly exchange that many bits per round to fix their strings. The second, $h(\Qx)$, is the number of bits sacrificed in privacy amplification. It is set by the phase error rate because the $Z$ and $X$ bases are complementary: an eavesdropper cannot gain information about the $Z$ basis without disturbing the $X$ basis. A low $\Qx$ therefore bounds what an eavesdropper could have learned, and hashing the key down by $h(\Qx)$ bits removes it.

Our pairwise baseline is the entanglement-based version of this protocol, in which a source distributes Bell pairs $|\Phi^+\rangle = (|00\rangle + |11\rangle)/\sqrt{2}$ and the two parties measure in randomly chosen bases. This is the E91-style implementation \cite{ekert1991}, and its asymptotic key rate is the same as prepare-and-measure BB84, since measuring one half of a maximally entangled state in a randomly chosen basis is equivalent to preparing a random state in that basis. We use the entanglement-based form because the GHZ protocol is entanglement-based, and putting both protocols in the same formalism means the same noise channels act on the same kind of object on both sides of the comparison.

\subsection{GHZ states and the N-BB84 protocol}\label{sec:nbb84}

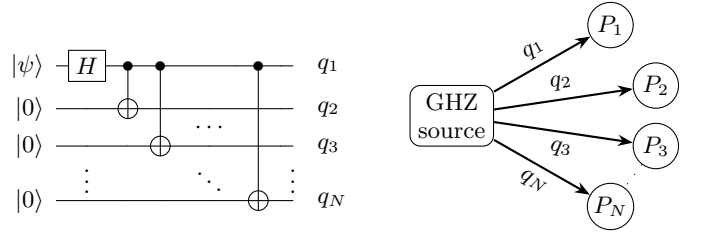
\begin{figure}[t]
  \centering

  \begin{minipage}[c]{0.50\linewidth}
    \centering
    \[\Qcircuit @C=0.5em @R=0.7em {
     \lstick{\ket{0}} &\gate{H} &\ctrl{1} &\ctrl{2} &\cds{4}{\cdots} &\ctrl{4} &\qw &\qw    &\rstick{q_1} \\
     \lstick{\ket{0}}    &\qw      &\targ    &\qw      &\qw             &\qw      &\qw &\qw    &\rstick{q_2} \\
     \lstick{\ket{0}}    &\qw      &\qw      &\targ    &\qw             &\qw      &\qw &\qw    &\rstick{q_3} \\
                         &\vdots   &         &         &\ddots          &         &    &\vdots & \\
     \lstick{\ket{0}}    &\qw      &\qw      &\qw      &\qw             &\targ    &\qw &\qw    &\rstick{q_N} \\
    }\]
  \end{minipage}
  \hfill
  \begin{minipage}[c]{0.40\linewidth}
    \centering
    \begin{tikzpicture}[
      party/.style={draw, circle, inner sep=1pt, minimum size=6mm},
      source/.style={draw, rounded corners, align=center},
      link/.style={-{Stealth[length=2mm]}, thick}
    ]
      \node[source] (src) at (0,0) {GHZ\\source};

      \node[party] (p1) at (2.1, 1.2) {$P_1$};
      \node[party] (p2) at (2.7, 0.4) {$P_2$};
      \node[party] (p3) at (2.7,-0.4) {$P_3$};
      \node[party] (pN) at (2.1,-1.2) {$P_N$};

      \draw[link] (src) -- node[above, sloped] {$q_1$} (p1);
      \draw[link] (src) -- node[above, sloped] {$q_2$} (p2);
      \draw[link] (src) -- node[below, sloped] {$q_3$} (p3);
      \draw[link] (src) -- node[below, sloped] {$q_N$} (pN);
      \draw[loosely dotted] (p3) -- (pN);
    \end{tikzpicture}
  \end{minipage}
  \caption{Preparation (left) and distribution (right) of an $N$-qubit GHZ state. The source prepares the state and sends one qubit $q_i$ to each party $P_i$, including the dealer $P_1$.}
  \label{fig:circuitghz}
\end{figure}

The $N$-party GHZ state is $|\mathrm{GHZ}_N\rangle = (|0\rangle^{\otimes N} + |1\rangle^{\otimes N})/\sqrt{2}$. It is the $+1$ eigenstate of the operators $X^{\otimes N}$ and $Z_i Z_j$ for each pair $(i, j)$, and these two groups of stabilizers are what the protocol measures.

The $N$-party generalization of BB84 (N-BB84) \cite{grasselli2018} is as follows. In each round, the parties share a GHZ state and all measure in the same basis, chosen from $\{Z, X\}$ according to a pre-shared key. Rounds in which everyone measures in $Z$ are key-generation rounds: with an ideal GHZ state, all $N$ outcomes are identical, so the dealer's bit is already the shared conference key bit. Rounds in which every party measures in $X$ are parameter-estimation rounds: on an ideal state, the parity of the $N$ outcomes is deterministic, so any deviation bounds how much phase information an eavesdropper can hold.

The two error rates that enter the key rate are therefore
\begin{align}
  \Qx     &= \frac{1 - \langle X^{\otimes N} \rangle}{2}, &
  Q_{1,i} &= \frac{1 - \langle Z_1 Z_i \rangle}{2},
  \label{eq:errrates}
\end{align}
where $\Qx$ is the error rate of the collective $X$ measurement and $Q_{1,i}$ is the pairwise $Z$-basis error rate between the dealer and receiver $i$. The asymptotic conference key rate is
\begin{equation}
  \rGHZ = 1 - h(\Qx) - \max_{2 \le i \le N} h(Q_{1,i}).
  \label{eq:ghzrate}
\end{equation}
The maximum appears because error correction must work for every receiver independently---the dealer cannot assume that the Bobs will collaborate---so the worst link sets the reconciliation cost.

\subsection{Noise channels}\label{sec:channels}

We model each arm of the network as an independent single-qubit channel acting on the qubit delivered to that party. We consider three channels.

Dephasing (phase damping) suppresses the off-diagonal elements of the density matrix without changing populations, with Kraus operators
\begin{equation}
  K_0 = \begin{pmatrix} 1 & 0 \\ 0 & \sqrt{1-\gamma} \end{pmatrix},
  \qquad
  K_1 = \begin{pmatrix} 0 & 0 \\ 0 & \sqrt{\gamma} \end{pmatrix}.
  \label{eq:dephasing}
\end{equation}
It is equivalent to a phase-flip channel applying a $Z$ gate with probability $p_z = (1 - \sqrt{1-\gamma})/2$.

The depolarizing channel applies each of the Pauli operators $X$, $Y$, and $Z$ with probability $\gamma/3$, and the identity with probability $1-\gamma$:
\begin{equation}
  \mathcal{E}(\rho) = (1-\gamma)\rho
    + \frac{\gamma}{3}\left(X\rho X + Y\rho Y + Z\rho Z\right).
  \label{eq:depol}
\end{equation}

Amplitude damping models energy loss to the environment, taking $|1\rangle$ toward $|0\rangle$:
\begin{equation}
  K_0 = \begin{pmatrix} 1 & 0 \\ 0 & \sqrt{1-\gamma} \end{pmatrix},
  \qquad
  K_1 = \begin{pmatrix} 0 & \sqrt{\gamma} \\ 0 & 0 \end{pmatrix}.
  \label{eq:ampdamp}
\end{equation}
The first two are Pauli channels, but amplitude damping is not. This difference matters in Section~\ref{sec:closedform}, where the closed-form results apply to the Pauli case, and in Section~\ref{sec:twirl}, where amplitude damping is handled by exact density-matrix simulation with a Pauli-twirled comparison.

\subsection{Structural asymmetry}\label{sec:asymmetry}

Why a crossover exists at all, and why it moves with channel type, is visible in the stabilizer structure.

Consider independent Pauli noise on each arm, with party $j$ subject to $X$, $Y$, and $Z$ errors with probabilities $p_j^x$, $p_j^y$, and $p_j^z$. Define the per-party aggregates $f_j = p_j^x + p_j^y$ (errors that anticommute with $Z$) and $g_j = p_j^z + p_j^y$ (errors that anticommute with $X$).

An $X$ error on any qubit commutes with $X^{\otimes N}$ and leaves it unchanged, while $Z$ and $Y$ errors anticommute with it and flip its sign. Because the sign flips are independent across parties, the expectation value is a product over all $N$ arms. In contrast, $Z_1 Z_i$ is flipped only by errors on qubits 1 and $i$, so its expectation value involves just two factors.

So $\Qx$ accumulates multiplicatively across the entire network while each $Q_{1,i}$ stays local to a pair. One noisy arm worsens the phase estimate for every party at once, whereas in the pairwise protocol both error types remain confined to the link on which they occur. This asymmetry is why the crossover tracks a channel's Pauli components, why admixing bit-flip error into dephasing is so damaging, why GHZ tolerance degrades as $N$ grows, and why a concentrated noise budget is worse than a spread one. The exact expressions for $\Qx$ and $Q_{1,i}$ are derived in Section~\ref{sec:closedform}.

\subsection{Resource counting}\label{sec:resource}

The two protocols do not consume network resources at an equal rate, so a comparison requires an explicit convention. In a bottleneck network, where all parties connect through a central station, the GHZ protocol distributes one state to all $N$ parties in a single network use, whereas the pairwise protocol requires $N-1$ separate links. We therefore compare $\rGHZ$ against $\rpair/(N-1)$, the convention used in the multipartite-advantage literature \cite{epping2017,memmen2023}.

We compare the asymptotic secret fractions per unit of network use. Transmission loss, detector efficiency, quantum memories, and finite-key corrections are outside the model, and their effect on the comparison is discussed in Section~\ref{sec:limitations}.

\subsection{Closed form for independent Pauli noise}\label{sec:closedform}

The stabilizer argument of Section~\ref{sec:asymmetry} carries through to an explicit rate. Let arm $j$ carry an independent Pauli channel applying $X$, $Y$, $Z$ with probabilities $p_j^x$, $p_j^y$, $p_j^z$, and recall the aggregates $f_j = p_j^x + p_j^y$ and $g_j = p_j^z + p_j^y$.

An error configuration flips the sign of $X^{\otimes N}$ if and only if it anticommutes with it on an odd number of arms. On arm $j$ this happens with probability $g_j$, independently of every other arm. For a single $\pm 1$-valued quantity that flips with probability $g$, the expectation is $(1-g)(+1) + g(-1) = 1 - 2g$, and independence means these factors multiply across all $N$ arms. The same argument applies to $Z_1 Z_i$, which is sensitive only to errors on arms 1 and $i$, occurring with probabilities $f_1$ and $f_i$, giving a two-factor product:
\begin{equation}
  \langle X^{\otimes N} \rangle = \prod_{j=1}^{N}(1 - 2g_j),
  \qquad
  \langle Z_1 Z_i \rangle = (1 - 2f_1)(1 - 2f_i).
  \label{eq:expectations}
\end{equation}
The error rates follow directly:
\begin{equation}
  \Qx = \tfrac{1}{2}\Bigl[1 - \prod_{j=1}^{N}(1 - 2g_j)\Bigr],
  \qquad
  Q_{1,i} = \tfrac{1}{2}\bigl[1 - (1 - 2f_1)(1 - 2f_i)\bigr].
  \label{eq:closedqs}
\end{equation}
For $0 \le f_j \le 1/2$ on every arm, $Q_{1,i}$ lies in $[0,1/2]$ and is nondecreasing in $f_i$. Since $h$ is also nondecreasing on this interval, the maximum over receivers is attained at the receiver with the largest $f_i$, giving
\begin{equation}
  \rGHZ = 1 - h(\Qx)
    - h\!\left(\tfrac{1}{2}\Bigl[1 - (1-2f_1)\bigl(1 - 2\max_{i\ge 2} f_i\bigr)\Bigr]\right).
  \label{eq:closedrate}
\end{equation}
The pairwise baseline on the same arms has, for the link $(1, i)$, $\Qx = \tfrac{1}{2}[1 - (1-2g_1)(1-2g_i)]$ and $\Qz = \tfrac{1}{2}[1 - (1-2f_1)(1-2f_i)]$; the conference rate is the minimum over receivers, divided by $N-1$.

We record this closed form because it does three things the simulation cannot: it establishes the comparison between complete concentration and an even split under the conditions of Section~\ref{sec:distribution}, and it gives the dealer-assignment rule of Section~\ref{sec:dealer} directly, with no search over permutations. It also provides an analytic check on the simulation at every parameter point and every group size, rather than only along the dephasing axis where $\Qz$ vanishes.

\section{Related Work}\label{sec:related}

The comparison between multipartite quantum communication and bipartite QKD has been studied from several perspectives, including network efficiency, noise tolerance, experimental feasibility, and security under realistic conditions. While GHZ-based conference key agreement (CKA) is often presented as a generalization of BB84, the literature consistently shows that its advantages are conditional, particularly in the presence of noise and imperfect channels.

Early work established the feasibility of multipartite quantum cryptography using GHZ correlations \cite{hillery1999}. Chen and Lo extended this to noisy environments, giving distillation schemes and prepare-and-measure constructions for conference key agreement and secret sharing, and computing the fidelity thresholds above which those protocols remain secure \cite{chen2007}. These results demonstrated that secure key generation survives degraded entanglement, but did not directly answer whether multipartite protocols are superior to bipartite ones.

That question was central to the work of Epping et al.~\cite{epping2017}, who showed that GHZ-based protocols can reduce the number of required network transmissions, especially in bottlenecked topologies. A single multipartite state can distribute correlations to all users simultaneously, while pairwise requires multiple independent links. Grasselli, Kampermann, and Bru{\ss} subsequently introduced the $N$-party BB84 protocol and proved its security in the finite-key regime \cite{grasselli2018}; the asymptotic rate derived there is the expression used throughout the present work. Murta et al.\ survey the resulting landscape \cite{murta2020}, including six-state, twin-field and single-photon-interference variants \cite{zhao2020,cao2021,grasselli2019}, device-independent protocols \cite{ribeiro2018,holz2020}, and anonymity-preserving extensions \cite{hahn2020,grasselli2022,webb2024}.

Despite this conceptual advantage, further work has shown that multipartite protocols are more sensitive to physical imperfections. Memmen, Eisert, and Walk show that the advantage of multipartite entanglement depends on channel loss, synchronization constraints, and network structure \cite{memmen2023}. The requirement that all qubits in a GHZ state arrive successfully in the same round leads to an exponential decrease in performance as the number of parties increases: in terms of the single-link transmission probability $p$, the multipartite yield scales as $p^N$ against $p^2/(N-1)$ for the bipartite scheme.

Two conventions from this literature carry over to the present comparison. The bipartite baseline is charged $N-1$ network uses against the GHZ protocol's one, a convention implemented in both of the above works by dividing the two-party rate by $N-1$. Rates are compared per network use rather than per distributed resource, since the latter penalizes the bipartite protocol for the fact that Bell pairs are easier to produce than GHZ states.

Experimental work supports this conclusion while also revealing additional details. Proietti et al.\ demonstrated GHZ-based CKA over tens of kilometers, confirming that multipartite protocols can be implemented in practice \cite{proietti2021}. Pickston et al.\ built a six-photon graph state supporting both four-party GHZ CKA and three-Bell-pair pairwise QKD on identical hardware, observing a higher rate for the GHZ protocol under controlled conditions \cite{pickston2023}. Both studies also show how much the advantage depends on well-controlled channels: in practice, unequal loss and noise across links introduce asymmetries that significantly degrade performance.

The asymmetry in question is where the experimental literature comes closest to the present work without resolving it. The supplementary material of Proietti et al.\ models a four-user star with an independent depolarizing parameter on each link and observes that the conference key rate depends on how noise is distributed across the network, identifying this as a degree of freedom with no two-party analogue. They frame the observation as an open problem, and the analysis is restricted to depolarizing channels. Pickston et al.\ encountered the same effect from the other direction: because noise was distributed unevenly across their six qubits, the achievable rate depended on which party was assigned the dealer role, and they optimized empirically over all permutations.

Underlying several of these observations is a structural difference in how noise propagates: errors in the GHZ protocol feed a global correlation term while pairwise errors stay confined to individual links, an asymmetry made precise in Section~\ref{sec:asymmetry}.

What the comparison literature has not varied is the type of noise. Epping et al.\ model each receiver's qubit under a depolarizing channel; Murta et al.\ plot the six-state and BB84 comparison for local depolarizing noise applied to the GHZ state; Memmen et al.\ use single-qubit depolarization together with time-dependent memory dephasing; the asymmetry analysis of Proietti et al.\ is depolarizing only; and device-independent analyses assume depolarized honest implementations \cite{ribeiro2018,holz2020}. Memory dephasing appears widely, but as a storage process rather than as a channel type whose effect on the crossover is compared against others. This uniformity has structural reasons: security proofs symmetrize arbitrary channels to a depolarized form, which erases distinctions between channel types by construction, and the parameter space grows quickly once each link carries its own multi-parameter channel family. The consequence is that the dependence of the crossover on channel type, as opposed to noise magnitude, party number, loss, or topology, has not been mapped quantitatively.

Amplitude damping is a particular casualty of this convention, since it is not a Pauli channel and therefore falls outside the Bell-diagonal and GHZ-diagonal machinery on which most CKA rate calculations rely. The standard workaround is the Pauli twirling approximation, which preserves the exact channel's average fidelity while discarding off-diagonal components of its Pauli transfer matrix \cite{tomita2014,katabarwa2015}. Because the accuracy of this approximation is not guaranteed once channels are composed across a network protocol, the present work reports both the exact simulation and the twirled comparison rather than assuming equivalence.

The present study builds on these findings by resolving the GHZ-versus-pairwise crossover along the axis the literature has left fixed. We map the crossover as a function of channel type rather than channel strength, and find that it depends strongly on the channel's bit- and phase-error contributions: channels producing only phase-type errors tolerate substantially more noise than those producing both, while the two channels producing both bit- and phase-type errors give crossovers agreeing to within 1.2\% despite differing Kraus structures. We derive closed-form key-rate expressions for independent per-party Pauli noise across $N$ parties, and use these expressions to establish an analytical comparison between concentrated and evenly distributed noise, extending the numerical observations of Proietti et al.\ for depolarizing channels. Prior work has analyzed large networks, experimental systems, loss, and finite-key effects; fewer studies isolate how noise composition and placement together determine the regime in which GHZ-based communication loses its advantage to simpler pairwise approaches.

\section{Methods}\label{sec:methods}

\subsection{Simulation framework}\label{sec:framework}

All results are produced by exact density-matrix simulation in Cirq. A single simulator instance, \texttt{cirq.DensityMatrixSimulator(dtype=np.complex128)}, was used. Error rates are obtained from expectation values computed directly from the density matrix rather than by sampling measurement outcomes, so the reported quantities have no shot noise, and the only numerical error is floating-point. Double precision is required throughout; the reason is quantified in Section~\ref{sec:validation}.

\subsection{State preparation and noise model}\label{sec:prep}

The $N$-party GHZ state is prepared by a Hadamard on the dealer's qubit followed by CNOT gates from the dealer to each of the $N-1$ receivers. The Bell pair for the pairwise baseline is the same construction at $N = 2$. After preparation, a single-qubit channel is applied independently to every qubit, including the dealer's, modeling the star topology in which the central station distributes one qubit down each arm. Three channels are used, each parameterized by a single strength $\gamma \in [0, 1]$: \texttt{cirq.phase\_damp}, \texttt{cirq.amplitude\_damp}, and \texttt{cirq.depolarize}, with Kraus operators as given in Section~\ref{sec:channels}.

\subsection{Error rate estimation}\label{sec:estimation}

From the noisy density matrix, we evaluate the stabilizer expectation values $\langle X^{\otimes N} \rangle$ and $\langle Z_1 Z_i \rangle$ for $i = 2, \ldots, N$ as Pauli-string observables, and convert them to the error rates $\Qx$ and $Q_{1,i}$ of Eq.~\eqref{eq:errrates}. The asymptotic conference key rate is Eq.~\eqref{eq:ghzrate}, floored at zero. The pairwise baseline applies the identical estimator to a two-qubit Bell state. A single link's error rates do not depend on the network size $N$; the pairwise side acquires its $N$-dependence only through the resource convention of Section~\ref{sec:convention}.

\subsection{Resource convention}\label{sec:convention}

Following Memmen et al.~\cite{memmen2023}, the pairwise rate is divided by $N-1$ before comparison, charging that protocol one network use per link against the GHZ protocol's single use for all parties. This is implemented directly in the pairwise rate function, so every comparison, threshold, and figure in this work uses $\rpair/(N-1)$. Section~\ref{sec:limitations} discusses the sensitivity of the thresholds to this choice.

\subsection{Locating the crossover}\label{sec:bisection}

Crossover thresholds $\gamma^*$ are found by bisection on the signed gap $\rGHZ(\gamma) - \rpair(\gamma)/(N-1)$ over $\gamma \in [0, 0.9]$, with tolerance $10^{-9}$, returning no result if the gap does not change sign on the interval. An analogous bisection locates the noise strength at which $\rGHZ$ first reaches zero. Bisection replaces the coarse grid search used in earlier iterations of this work: on a 0.005 grid, thresholds are accurate only to the grid spacing, which turned out to be comparable to the very differences between channels we were trying to resolve in Section~\ref{sec:composition}.

\subsection{Infidelity normalization}\label{sec:infidelity}

Cirq parameterizes the three channels independently, so a given $\gamma$ does not denote the same physical disturbance across channels, and raw thresholds are not comparable. Each threshold is therefore converted to the average channel infidelity $1 - \Favg$, computed by applying the channel's Kraus operators to each of the six axis states $|0\rangle$, $|1\rangle$, $|\pm\rangle$, $|{\pm i}\rangle$, evaluating the fidelity to the input state, and averaging. These six states form a 2-design, so the average reproduces the Haar average over pure input states.

\subsection{Stacked and asymmetric noise}\label{sec:variants}

Two variants of the circuit constructors support the remaining results. Stacked-noise versions accept independent strengths for the amplitude-damping, phase-damping, and depolarizing channels and apply all selected channels in sequence on each qubit. The channels are applied in the order amplitude damping, phase damping, and depolarizing noise. These produce the admixture sweep of Section~\ref{sec:composition}, in which the crossover threshold is expressed in terms of the base dephasing-strength parameter $\gamma$. An asymmetric variant accepts a per-arm list of noise strengths, used for the fixed-budget comparisons of Section~\ref{sec:distribution} and the dealer-permutation test of Section~\ref{sec:dealer}.

\subsection{Data and code availability}\label{sec:code}

All code is available at \url{https://github.com/QHSRP/Quantum-Computing-BB84}. The rate estimators, circuit constructors (including the stacked-noise and asymmetric variants of Section~\ref{sec:variants}), crossover routines, and infidelity calculation are in \texttt{GHZ\_CKA.py}.

\section{Results}\label{sec:results}

\subsection{Validation against closed forms}\label{sec:validation}

\begin{figure}[t]
  \centering
  \includegraphics[width=\columnwidth]{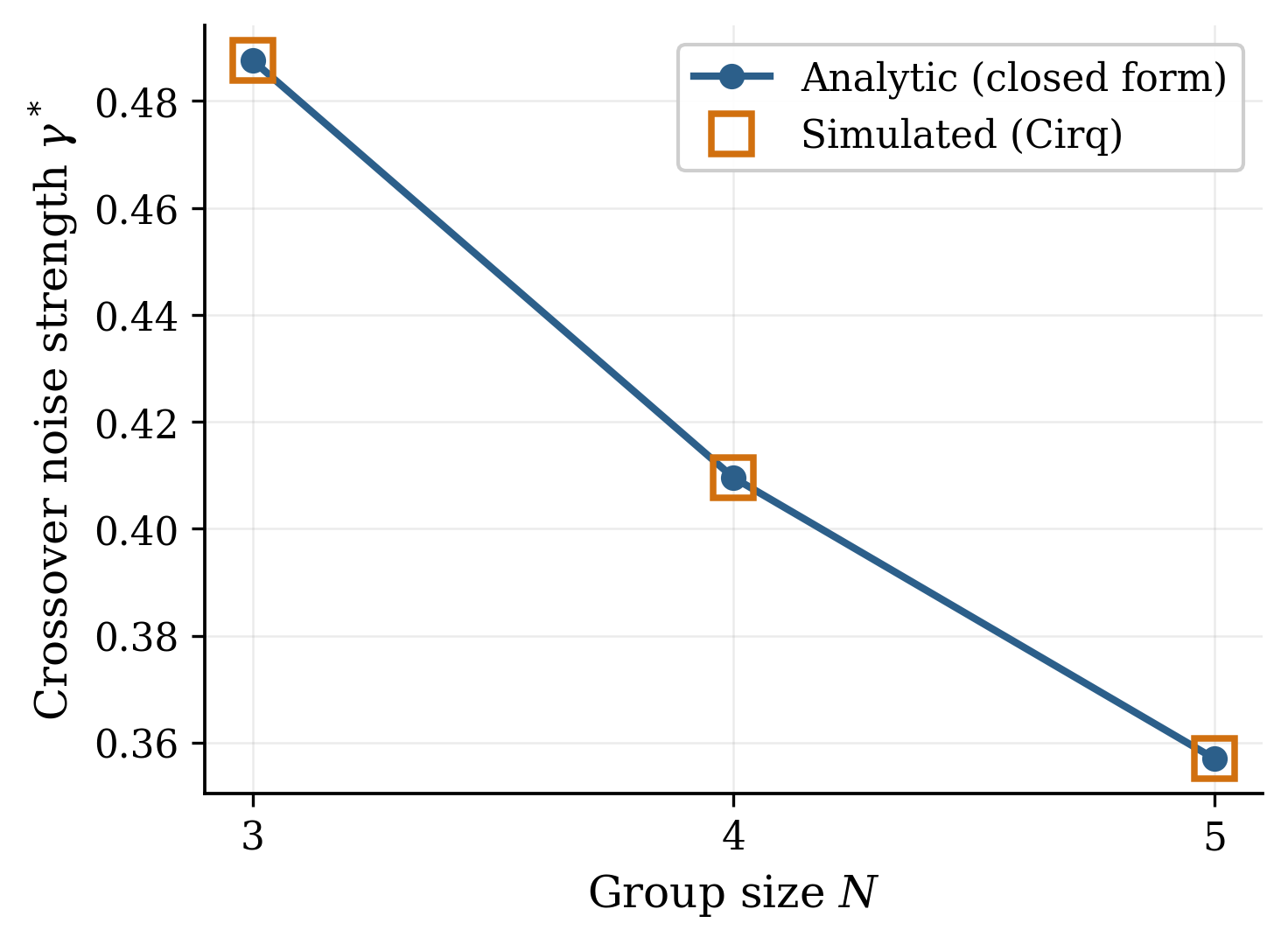}
  \caption{Dephasing crossover threshold $\gamma^*$ versus group size:
  closed form (line) against independent Cirq simulation (open
  squares), $N = 3$...$5$. The two bisections agree to
  $2.6\times10^{-10}$, limited by the solver tolerance; the maximum
  pointwise key-rate discrepancy across 46 noise values and all three
  group sizes is $1.2\times10^{-15}$.}
  \label{fig:validation}
\end{figure}

Under pure dephasing, the key rates admit an exact closed form. Phase damping produces no bit-flip errors, so $\Qz = 0$ identically and the GHZ rate reduces to a single entropy term, $\rGHZ = 1 - h(\Qx)$ with $\Qx = (1 - (1-\gamma)^{N/2})/2$, while the Bell pair sees $\Qx = \gamma/2$ and $\Qz = 0$. These expressions give an independent check on the entire simulation, from circuit construction to expectation values to the rate formula.

Across 46 noise values spanning $\gamma \in [0, 0.9]$ and group sizes $N = 3, 4, 5$, the largest discrepancy between simulated and analytic key rate is $\max |r_{\mathrm{sim}} - r_{\mathrm{analytic}}| = 1.2\times10^{-15}$. Solving for the crossover location on each side independently, the two agree to $2.6\times10^{-10}$ (Fig.~\ref{fig:validation}), a figure set by the bisection tolerance rather than by simulation error.

This level of agreement requires double precision. Cirq's \texttt{DensityMatrixSimulator} defaults to \texttt{complex64}, which introduces errors of order $10^{-6}$ in the key rate near the crossover: $h(x)$ has a divergent derivative at $x = 0$ and $1$, so single-precision error in $\Qx$ is amplified exactly where the two protocols are closest, which is where the crossover sits. All results below use \texttt{dtype = np.complex128}.

\subsection{Dependence of the crossover on noise composition}\label{sec:composition}

\begin{figure*}[t]
  \centering
  \includegraphics[width=\textwidth]{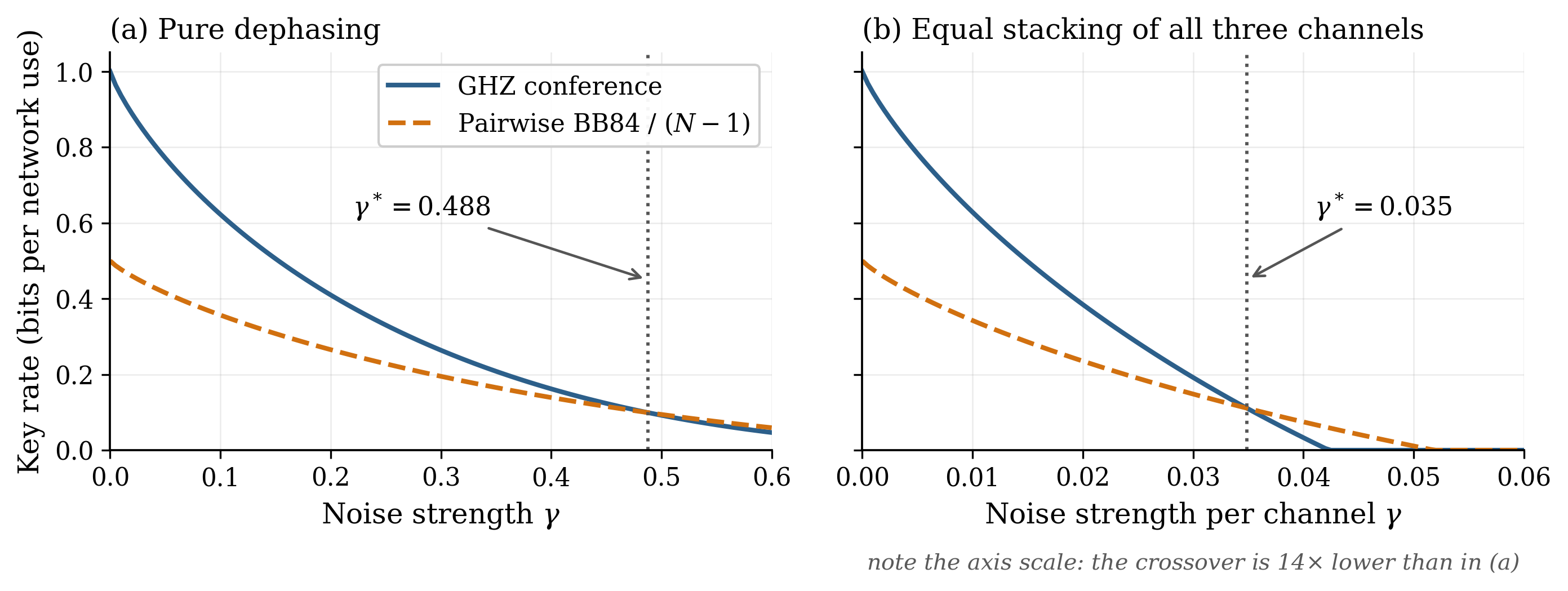}
  \caption{The GHZ--pairwise crossover at two noise compositions,
  $N = 3$. (a)~Pure dephasing: the protocols cross at
  $\gamma^* = 0.488$. (b)~Equal stacking of dephasing, amplitude
  damping, and depolarizing at a common strength: $\gamma^* = 0.035$, a
  factor of 14 lower. Note the tenfold difference in axis scale between
  panels, although the protocols and resource convention are
  identical.}
  \label{fig:composition}
\end{figure*}

Each protocol was simulated across each of the three channels and the crossover located by performing bisection on $\rGHZ - \rpair/(N-1)$ (Fig.~\ref{fig:composition} shows the two extremes). Table~\ref{tab:crossovers} collects the results.

The table also reports $\gamma_{\mathrm{dies}}$, the noise strength at which the GHZ rate itself reaches zero. At every crossover, the GHZ protocol is still producing key; $\gamma^*$ precedes $\gamma_{\mathrm{dies}}$ by 12--17\% for the channels that have one.
The crossover is an economic boundary set by the resource convention, not a security failure of the multipartite protocol. Under dephasing, the GHZ rate remains positive over the entire range examined.

\begin{table}[t]
  \caption{Crossover noise strength $\gamma^*$, the strength
  $\gamma_{\mathrm{dies}}$ at which the GHZ rate reaches zero (dashes:
  no zero on $\gamma \in [0, 0.9]$), and the crossover expressed as
  average channel infidelity $1 - \Favg$.}
  \label{tab:crossovers}
  \begin{ruledtabular}
  \begin{tabular}{llccc}
  Channel & $N$ & $\gamma^*$ & $\gamma_{\mathrm{dies}}$ & $1-\Favg$ at $\gamma^*$ \\
  \colrule
  Dephasing         & 3 & 0.48759 & ---     & 0.09472 \\
                    & 4 & 0.40956 & ---     & 0.07720 \\
                    & 5 & 0.35701 & ---     & 0.06604 \\
  \colrule
  Amplitude         & 3 & 0.11739 & 0.13920 & 0.03974 \\
  damping           & 4 & 0.10462 & 0.12105 & 0.03536 \\
                    & 5 & 0.09512 & 0.10822 & 0.03210 \\
  \colrule
  Depolarizing      & 3 & 0.05963 & 0.07197 & 0.03975 \\
                    & 4 & 0.05269 & 0.06209 & 0.03513 \\
                    & 5 & 0.04758 & 0.05513 & 0.03172 \\
  \end{tabular}
  \end{ruledtabular}
\end{table}

In raw parameter terms, the thresholds differ by a factor of 8.18 at $N = 3$. However, Cirq parameterizes \texttt{phase\_damp}, \texttt{amplitude\_damp}, and \texttt{depolarize} independently, so $\gamma$ carries a different meaning in each case, and the comparison is between incommensurable quantities. We therefore convert each threshold to the average infidelity of the underlying single-qubit channel, averaged over the six axis states, which form a 2-design and reproduce the Haar average. The rightmost column of Table~\ref{tab:crossovers} is the comparable one.

Normalized, the spread shrinks to a factor of 2.38 at $N = 3$ (2.20 at $N = 4$, 2.08 at $N = 5$). Most of the apparent order-of-magnitude effect was parameterization.

Amplitude damping and depolarizing land almost on top of each other. At $N = 3$, their thresholds are 0.03974 and 0.03975, agreeing to better than one part in a thousand, and they agree to within 1.2\% at every group size tested, even though they differ by a factor of two in raw $\gamma$ and have entirely different Kraus structure.
The channels do not sort by name, but by whether they produce bit-flip-type errors as well as phase-type ones.

This is exactly what the stabilizer structure of Section~\ref{sec:asymmetry} predicts. The GHZ rate has two entropy terms: $h(\Qx)$, made from $\langle X^{\otimes N} \rangle$ and therefore accumulating multiplicatively across all $N$ arms, and $\max_i h(Q_{1,i})$, built from $\langle Z_1 Z_i \rangle$ and confined to a single pair. Dephasing produces only phase-type errors, so $\Qz = 0$ exactly and the GHZ protocol pays nothing on the second term; its tolerance is correspondingly higher, by a factor of roughly 2 in infidelity (2.4 at $N = 3$, narrowing to 2.1 at $N = 5$). Amplitude damping and depolarizing both feed both terms, and once matched in infidelity, their crossovers nearly coincide.

\begin{figure}[t]
  \centering
  \includegraphics[width=\columnwidth]{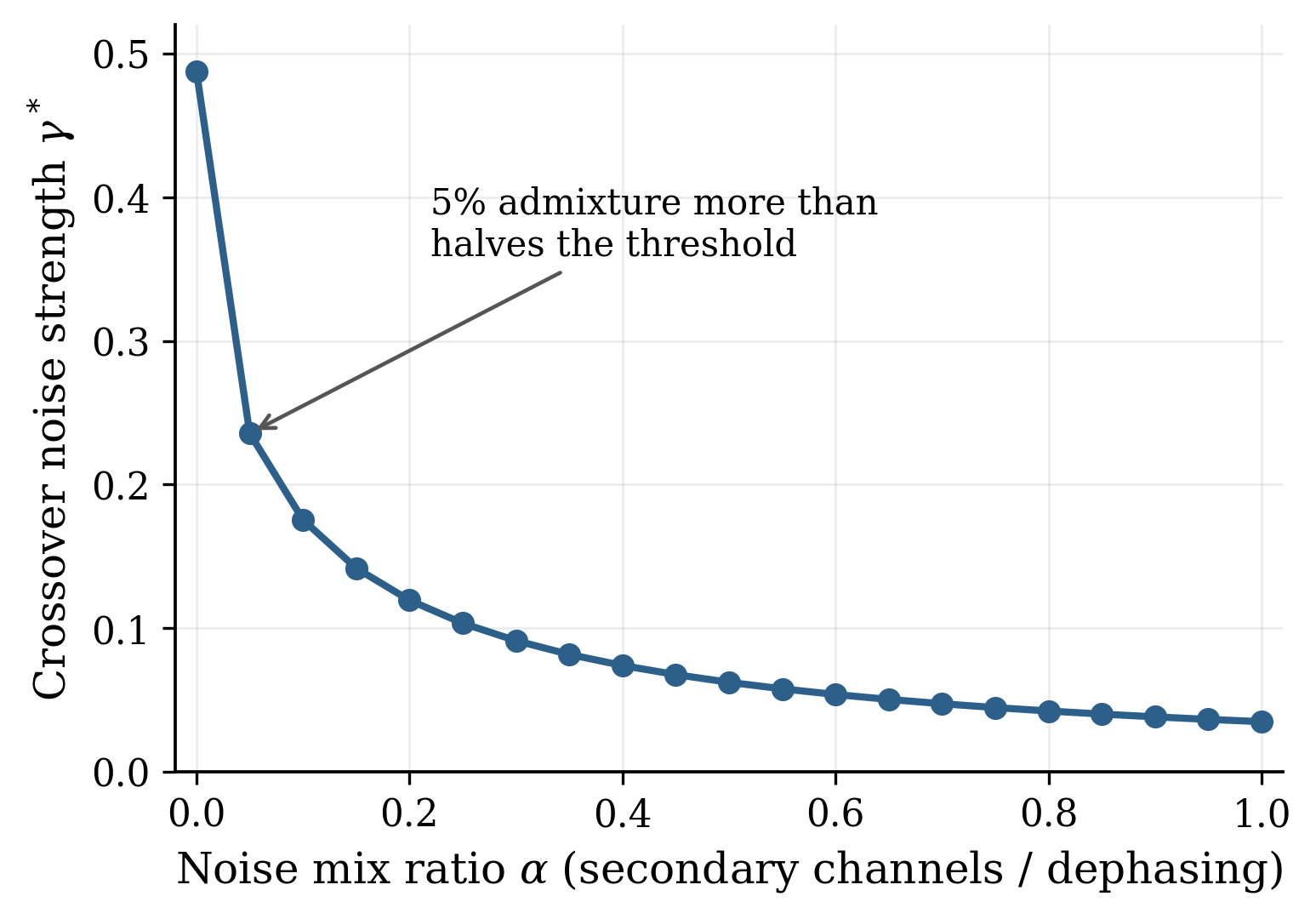}
  \caption{Crossover threshold $\gamma^*$ as bit-flip-type error is
  admixed into pure dephasing at ratio $\alpha$ (amplitude damping and
  depolarizing each at strength $\alpha\gamma$), $N = 3$. The thresholds are expressed in terms of the base dephasing-strength parameter $\gamma$.
  The collapse is front-loaded: 5\% admixture more than halves the
  threshold, because it activates the $h(\Qz)$ term that pure dephasing
  leaves identically zero.}
  \label{fig:mix}
\end{figure}

A cleaner demonstration of the same effect avoids the cross-channel comparison entirely. We stack a dephasing channel with amplitude-damping and depolarizing channels at strength $\alpha$ times the dephasing strength and sweep $\alpha$ from 0 to 1; every point in the sweep uses the same parameterization, allowing comparison of the crossover in the base dephasing-strength parameter as the admixture ratio varies (Fig.~\ref{fig:mix}). The threshold falls from 0.48759 at $\alpha = 0$ to 0.03488 at $\alpha = 1$, a factor of 14, and the collapse is front-loaded: admixing secondary channels at just 5\% of the dephasing strength cuts the crossover threshold $\gamma^*$ by more than half. The GHZ advantage is disproportionately sensitive to bit-flip-type errors because they activate an entropy term that pure dephasing leaves identically zero.

\subsection{Scaling with group size}\label{sec:scaling}

\begin{figure*}[t]
  \centering
  \includegraphics[width=\textwidth]{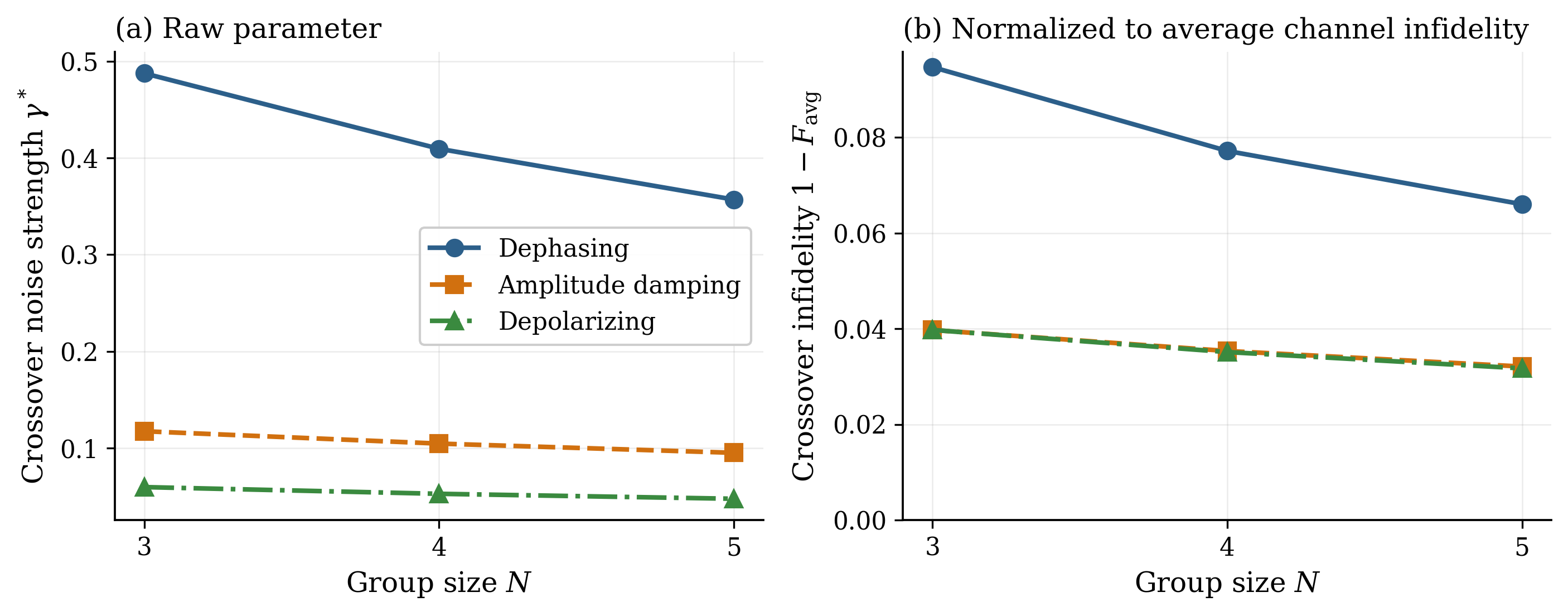}
  \caption{Crossover thresholds versus group size for the three
  channels. (a)~Raw parameter $\gamma^*$, which is not comparable
  across channels because Cirq parameterizes each independently.
  (b)~The same thresholds normalized to average channel infidelity
  $1-\Favg$: amplitude damping and depolarizing agree to within
  1.2\% at every $N$, while dephasing sits a factor of 2.4 to 2.1 higher.
  The channel ordering in (a) is an artifact of parameterization; the
  structure in (b) is physical.}
  \label{fig:scaling}
\end{figure*}

Crossover thresholds fall monotonically with $N$ for every channel, in both raw and infidelity-normalized terms (Fig.~\ref{fig:scaling}). For dephasing, the normalized threshold falls from 0.09472 at $N = 3$ to 0.0660 at $N = 5$; for depolarizing, it falls from 0.0398 to 0.0317. The individual pairwise link rate is independent
of $N$, although the resource-normalized
conference rate decreases through the
$1/(N-1)$ prefactor. The observed crossover
shift indicates that the accumulation of
phase errors across additional GHZ arms
outweighs this resource advantage.

The direction of the effect is not new. Epping et al.\ give threshold QBERs as a function of party number \cite{epping2017}, and Memmen et al.\ derive explicit depolarizing thresholds against $N$ for essentially this comparison \cite{memmen2023}. Our contribution on this specific point is confirmatory, and is reported mainly because it is a necessary precondition for reading Table~\ref{tab:crossovers} correctly.

The noise model does, however, differ significantly from that of Epping et al. Our channels act independently on each qubit after distribution, so the growth in $\Qx$ with $N$ reflects the accumulation of local errors across an increasing number of arms, whereas Epping's threshold analysis applies depolarization at the level of the distributed state. The two are not in conflict and should not be read as competing measurements of the same quantity. Nor is this superdecoherence, the faster-than-linear decay that occurs when qubits couple to a common reservoir \cite{kam2024}; our noise is uncorrelated across arms, so the correct description is local noise accumulation.

\subsection{Concentrated noise is worse than distributed noise}\label{sec:distribution}

\begin{figure}[t]
  \centering
  \includegraphics[width=\columnwidth]{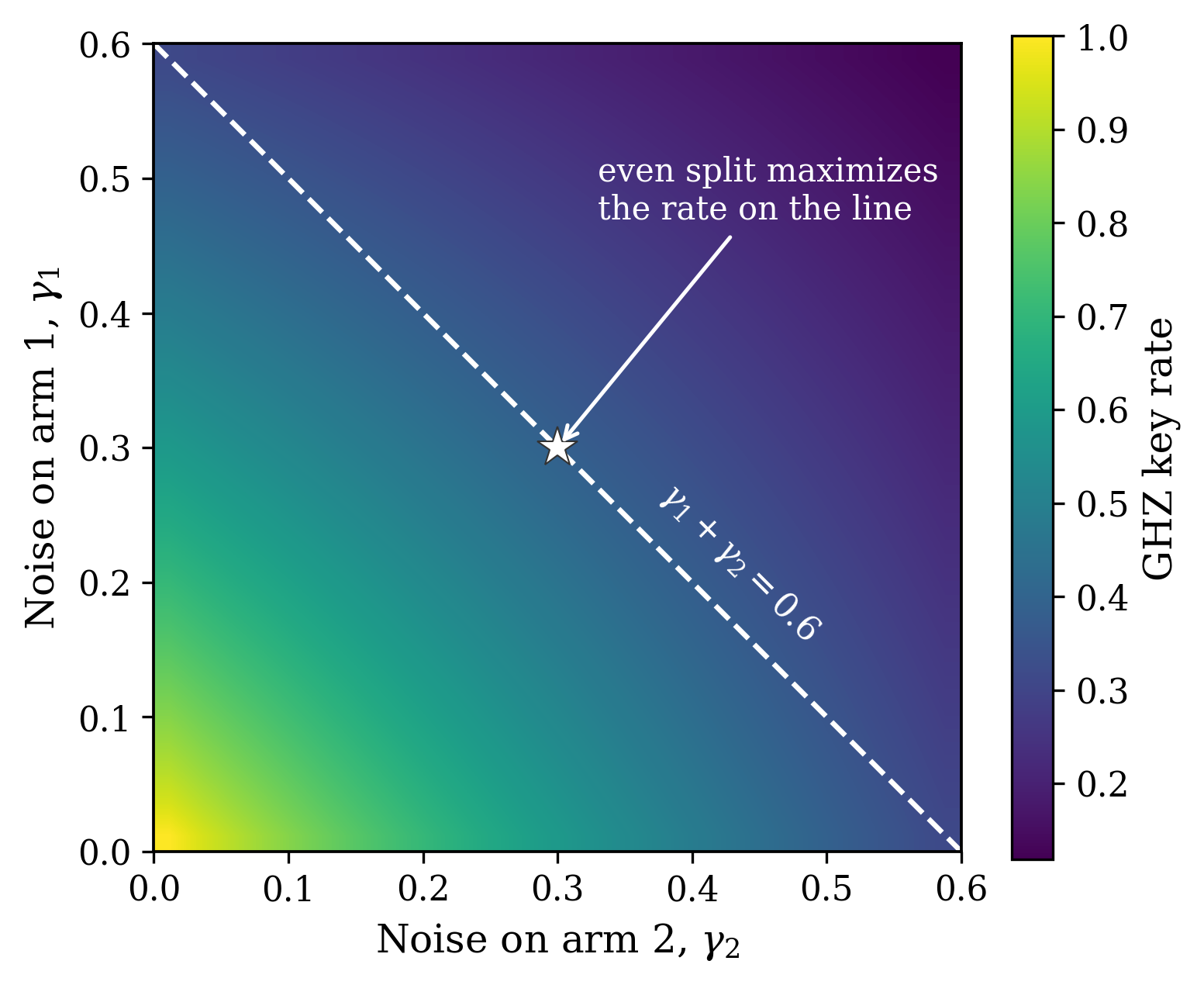}
  \caption{GHZ key rate under asymmetric dephasing on arms 1 and 2,
  with arm 3 noiseless, $N = 3$. The dashed line marks a fixed total
  budget $\gamma_1 + \gamma_2 = 0.6$; along it the rate is maximized at
  the even split (star), the arithmetic--geometric-mean result proved
  in Section~\ref{sec:distribution}, and falls monotonically toward
  either concentrated end.}
  \label{fig:heatmap}
\end{figure}

Holding the total noise budget $S = \sum_j \gamma_j$ fixed, we compare an even split across all arms against concentration on a single arm (Fig.~\ref{fig:heatmap}; Table~\ref{tab:budget}).

\begin{table}[t]
  \caption{GHZ key rate under even versus concentrated noise at fixed
  budget $S$, $N = 3$.}
  \label{tab:budget}
  \begin{ruledtabular}
  \begin{tabular}{lccccc}
  Channel & $S$ & Even split & Concentrated & Loss & \% drop \\
  \colrule
  Dephasing         & 0.30 & 0.6226 & 0.5920 & 0.0307 & 4.9   \\
                    & 0.60 & 0.4100 & 0.3117 & 0.0982 & 24.0  \\
                    & 0.90 & 0.2640 & 0.0734 & 0.1906 & 72.2  \\
  \colrule
  Amplitude         & 0.15 & 0.4958 & 0.3779 & 0.1179 & 23.8  \\
  damping           & 0.30 & 0.1862 & 0.0000 & 0.1862 & 100.0 \\
  \colrule
  Depolarizing      & 0.09 & 0.4434 & 0.3451 & 0.0983 & 22.2  \\
  \end{tabular}
  \end{ruledtabular}
\end{table}

The penalty appears for every channel tested, and grows with the total budget. It vanishes as $S \to 0$, where both configurations approach the noiseless rate, and reaches total suppression at large $S$, where concentration alone can drive the rate to zero while the even split still yields key. Sampling 4000 random splits of a fixed budget $S = 0.9$ under dephasing from a Dirichlet distribution, the maximum rate occurs at the near-equal split and the minimum at the most concentrated draw.

For independent Pauli channels, an analytical
comparison is possible at fixed aggregate budgets
$S_f = \sum_j f_j$ and $S_g = \sum_j g_j$.
Consider $0 \le S_f,S_g \le 1/2$, and compare
an equal allocation of both budgets across all arms
with their complete concentration on one arm.

By the arithmetic--geometric mean inequality,
\begin{equation}
  \prod_{j=1}^{N}(1-2g_j)
  \leq \left(1-\frac{2S_g}{N}\right)^N.
\end{equation}
Equality is attained at the even split, which
therefore minimizes the phase-error contribution.

For the bit-error contribution, complete
concentration gives $\max_i Q_{1,i}=S_f$, whereas
the even split gives
\begin{equation}
  \max_i Q_{1,i}
  = \frac{2S_f}{N}\left(1-\frac{S_f}{N}\right)
  \leq S_f.
\end{equation}
Since the binary entropy is nondecreasing on
$[0,1/2]$, neither entropy cost is larger for
the even split. Hence the even split yields
a key rate at least as large as complete
concentration under these conditions.

This comparison does not imply that an equal
allocation maximizes the full key rate over
all possible asymmetric noise distributions.

For the dephasing parameterization used in our
simulations, $\langle X^{\otimes N}\rangle
= \sqrt{\prod_j(1-\gamma_j)}$.
At fixed $S=\sum_j\gamma_j$, the
arithmetic--geometric mean inequality therefore
also establishes that an even split maximizes
the dephasing-only key rate.

This answers a question posed but not resolved by Proietti et al.~\cite{proietti2021}, whose supplementary material observes numerically that conference key rates depend nontrivially on how noise is distributed across a star network and identifies this as a degree of freedom with no two-party analogue. Their analysis is restricted to depolarizing channels; the comparison above extends to independent Pauli channels under the stated aggregate-budget conditions and follows analytically from the closed form of Section~\ref{sec:closedform}. Pickston et al.\ saw the same effect experimentally \cite{pickston2023}, finding that the achievable rate depended on which party was assigned the dealer role because noise was distributed unevenly across their six-qubit resource state.

\subsection{Amplitude damping: exact simulation versus Pauli twirling}\label{sec:twirl}

The closed form of Section~\ref{sec:closedform} assumes Pauli noise, and amplitude damping is not a Pauli channel, so we do not apply it directly. The standard workaround is the Pauli twirling approximation \cite{tomita2014,katabarwa2015}, which replaces the channel with its Pauli-diagonal part: $p_x = p_y = \gamma/4$ and $p_z = (1 - \sqrt{1-\gamma})/2 - \gamma/4$. Instead of assuming this approximation is adequate, we report both.

\begin{table}[t]
  \caption{GHZ key rate under exact and twirled amplitude damping,
  $N = 3$.}
  \label{tab:twirl}
  \begin{ruledtabular}
  \begin{tabular}{cccc}
  $\gamma$ & Exact & Twirled & Difference \\
  \colrule
  0.020  & 0.74891 & 0.74778 & 0.00113 \\
  0.050  & 0.49583 & 0.49044 & 0.00538 \\
  0.080  & 0.29819 & 0.28660 & 0.01159 \\
  0.100  & 0.18617 & 0.16970 & 0.01647 \\
  0.1174 & 0.09867 & 0.07758 & 0.02109 \\
  \end{tabular}
  \end{ruledtabular}
\end{table}

Pauli twirling preserves
$\langle X^{\otimes N}\rangle$ exactly for
amplitude damping, so the difference between
the exact and twirled key rates arises entirely
from the pairwise bit-error rates.

In the Heisenberg picture, the exact
amplitude-damping channel acts as
\begin{equation}
  \mathcal{E}^{\dagger}(Z)
  = (1-\gamma)Z+\gamma I,
\end{equation}
whereas its Pauli twirl removes the identity
contribution. For two arms with equal damping
strength, this gives
\begin{align}
  Q_z^{\mathrm{exact}} &= \gamma-\gamma^2,\\
  Q_z^{\mathrm{twirled}} &= \gamma-\frac{\gamma^2}{2}.
\end{align}
The twirled channel therefore overestimates the
pairwise bit-error rate by exactly $\gamma^2/2$.
This discrepancy arises from the nonunital
component of amplitude damping, which is removed
by Pauli twirling. Propagated to the threshold itself---twirling both protocols, since in practice the approximation would be applied throughout---the crossover lands at 0.11358 against the exact 0.11739, a 3.2\% error at $N = 3$, falling to 2.4\% at $N = 5$.

This approximation is therefore adequate for qualitative statements and for placing amplitude damping in Table~\ref{tab:crossovers}, but not for precise quoted threshold values. All amplitude-damping figures in this work are exact density-matrix simulations; the twirled values appear only in this comparison. This pessimistic bias of the twirl is consistent with what Tomita and Svore observed for surface-code thresholds under the same approximation \cite{tomita2014}, here isolated to a single term of the key rate with its coefficient identified exactly.

\section{Discussion}\label{sec:discussion}

\subsection{One mechanism accounts for all of the results}\label{sec:mechanism}

The four effects reported above are not independent findings. They are four faces of one structural fact about the N-BB84 rate.

The GHZ protocol pays two entropy costs. The first, $h(\Qx)$, derives from $\langle X^{\otimes N} \rangle$, whose expectation value under independent per-arm noise is a product over all $N$ arms. The second, $\max_i h(Q_{1,i})$, derives from $\langle Z_1 Z_i \rangle$, which involves two arms at a time. The pairwise protocol restricts both costs to a single link. Every result follows from asking what a given disturbance does to that product.

\paragraph{Channel type.} A channel that produces no bit-flip-type errors leaves $\Qz$ identically zero, so the GHZ protocol pays only the global term. Dephasing does this exactly, and tolerates roughly twice the infidelity before losing to the pairwise baseline. Amplitude damping and depolarizing both feed both terms, and when matched in infidelity, their thresholds agree to within 1.2\%.
For independent Pauli channels, the key rates are determined by the per-arm bit- and phase-error aggregates. The amplitude-damping comparison suggests that these aggregates remain informative beyond the Pauli-channel family.

\paragraph{Admixture.} The collapse in Fig.~\ref{fig:mix} is front-loaded because it reflects the activation of a term rather than its growth. Going from pure dephasing to 5\% admixture more than halves the crossover threshold $\gamma^*$, because $h(\Qz)$ moves from identically zero to small, and the marginal cost of the first bit-flip error is much larger than that of any subsequent one.

\paragraph{Group size.} Adding a party adds a factor to the product in $\Qx$ while leaving each $Q_{1,i}$ a two-factor quantity. The GHZ rate degrades with additional arms, while the individual pairwise link rate remains unchanged. Under our resource convention, the net effect is a downward shift in the observed crossover.

\paragraph{Noise distribution.}
At fixed total phase-error budget, the product
$\prod_j (1-2g_j)$ is maximized by an even
allocation in the nonnegative-factor regime.
For the full key rate, the comparison in
Section~\ref{sec:distribution} shows that
complete concentration cannot outperform
an even allocation of the same aggregate
bit- and phase-error budgets under the
stated conditions.

That all four predictions hold in simulation, on top of the machine-precision agreement of Section~\ref{sec:validation}, is what convinces us the mechanism is correctly identified, not merely consistent with one dataset.

\subsection{A practical corollary: which party should hold the key}\label{sec:dealer}

The closed form has an immediate operational consequence that we did not anticipate. The dealer's arm noise enters every pairwise term $Q_{1,i}$, while each receiver's noise enters only one. For independent Pauli noise in the regime $0 \le f_j \le 1/2$, the dealer role can therefore be optimally assigned to a party with the smallest bit-error aggregate $f_j=p_j^x+p_j^y$.

Simulating all six assignments of the three noise strengths to the three arms for $N = 3$, with arm noise 0.02, 0.06, and 0.12 under amplitude damping, confirms this. $\Qx$ is identical across all six permutations, since it is a symmetric product over arms. The key rate is 0.3575 when the cleanest arm holds the dealer role and 0.3024 otherwise, an 18\% improvement for a choice that costs nothing.

The degenerate cases are informative too. Assigning the dealer role to the second-cleanest or noisiest arm gives identical rates, because in both cases the maximum over receivers selects the pair formed by the two noisiest arms. The rate is set by the two worst arms unless the dealer is the cleanest one, in which case it is set by the best and worst.

The prediction comes with its own test: the effect operates entirely through $\Qz$, so it should vanish for channels that do not produce bit-flip errors. Repeating the same six-permutation test under dephasing gives a rate of 0.714377 for every assignment, identical to six decimal places.

This explains an observation by Pickston et al.~\cite{pickston2023}, who found that the achievable rate in their six-photon experiment depended on which party held the dealer role, and optimized empirically over all permutations. The closed form identifies why the optimization helps, predicts which permutation wins without a search, and predicts the conditions under which it will not help at all.

\subsection{Relation to prior work}\label{sec:priorwork}

Two of the four results have partial precedent, so we state exactly which parts are new.

The movement of the crossover with party number is not. Epping et al.\ give threshold error rates as a function of $N$ \cite{epping2017}, and Memmen et al.\ derive explicit depolarizing thresholds against $N$ for essentially the comparison studied here \cite{memmen2023}. Section~\ref{sec:scaling} confirms these under a per-qubit noise model rather than establishing them, and is included because the $N$-dependence is a precondition for reading Table~\ref{tab:crossovers} correctly, not as a claim of priority. What the type-resolved analysis adds is that the thresholds these works report are specific to depolarizing channels in a way not visible from within a single-channel study: a fidelity-matched comparison places depolarizing and amplitude damping within 1.2\% of each other and dephasing a factor of 2.1--2.4 away, so a depolarizing threshold generalizes to some channels and badly misestimates others.

The noise-distribution result answers a question that was posed rather than resolved. Proietti et al.\ observe numerically that conference key rates in a four-user star depend on how depolarizing noise is allocated across links \cite{proietti2021}, and identify this as a degree of freedom with no two-party analogue. Section~\ref{sec:distribution} establishes a comparison between complete concentration and an even split for independent Pauli noise under fixed aggregate budgets, and derives it analytically from the arithmetic--geometric mean inequality rather than mapping it numerically. The dealer-assignment corollary of Section~\ref{sec:dealer} is the same mechanism applied to a choice that costs nothing to make, and it accounts for the permutation optimization Pickston et al.\ arrived at empirically \cite{pickston2023}, including the prediction that the optimization returns nothing under pure dephasing.

Two clarifications prevent apparent conflicts with prior work. Our channels act independently on each qubit after distribution, whereas Epping's threshold analysis applies depolarization at the level of the distributed state; these are different models, and the resulting thresholds are not competing estimates of one quantity. Separately, the growth of GHZ fragility with $N$ is local noise accumulation, not superdecoherence (Section~\ref{sec:scaling}); our noise is uncorrelated across arms, so linear-in-$N$ accumulation is the expected behavior for this model, not evidence of collective coupling.

\subsection{Limitations}\label{sec:limitations}

Transmission loss is not modeled. This is the most important omission. Loss affects the GHZ protocol disproportionately, since all $N$ arms must succeed simultaneously, and Memmen et al.\ show that including it removes the multipartite advantage entirely for links beyond roughly 15~km of direct-transmission fiber \cite{memmen2023}, in a network where the noise-only analysis would predict an advantage. Our thresholds should therefore be read as characterizing only the noise-induced component of the crossover, and as upper bounds on where the boundary sits in a lossy deployment.

The analysis is asymptotic. No finite-key corrections are applied \cite{grasselli2018,li2021}, and these would not affect the two protocols equally: the GHZ protocol requires all parties to measure in a matching basis, reducing usable rounds, while the pairwise protocol must compose $N-1$ keys to a common security parameter. We do not claim these effects cancel, only that we have not quantified either.

The resource-counting convention is assumed. We charge the pairwise protocol $N-1$ network uses, following the bottleneck-network convention. All threshold values scale with this choice, and a non-bottleneck topology would remove the prefactor entirely.

The model is at qubit level. Source imperfections, multi-pair emission, detector dark counts, and basis misalignment are absent. In particular, photon-number-splitting attacks are not structurally testable with a QBER-based qubit model, since they can occur without introducing detectable error, so nothing here speaks to implementation security.

Group sizes are small. Density-matrix simulation costs memory of order $4^N$, so we restrict direct simulation to $N \le 5$. The trend is established over three points, and extrapolation beyond that range rests on the closed form rather than on simulation.

\subsection{Outlook}\label{sec:outlook}

The most valuable extension is the one our limitations imply. Combining the type-resolved analysis here with the loss-and-memory modeling of Memmen et al.~\cite{memmen2023,memmen2026} would produce an advantage boundary in both noise composition and distance, which is the object a network designer needs and which neither analysis alone provides.

One smaller question arose along the way and was answered. Amplitude damping and depolarizing agree to 0.03\% at $N = 3$ but separate slowly with group size, reaching 1.2\% at $N = 5$. The twirl analysis of Section~\ref{sec:twirl}
isolates a non-Pauli contribution to this
comparison: the exact amplitude-damping channel
reduces the pairwise bit-error rate by
$\gamma^2/2$ relative to its Pauli twirl for
equal-strength damping on both arms. Separately, the closed form extends to arbitrary $N$ at negligible cost, so the $N$-dependence could be pushed well past the simulation limit, subject to spot-checking at the largest tractable size.

Finally, the mechanism identified here suggests
an experimental target. For independent Pauli
channels with identical per-arm bit- and
phase-error aggregates, our closed-form
expressions predict identical crossovers.
Testing this prediction for physically distinct
noise processes would also help quantify the
importance of non-Pauli contributions.

\begin{acknowledgments}
This work was carried out as part of the QHSRP at Frontier Technology Institute. The authors thank the Frontier Technology Institute for providing the opportunity and support to conduct this research. We thank our teaching assistant, Aaditya Bhattacharya, for technical guidance throughout the research project, and our instructor, Kian Jansepar, for teaching the quantum computing coursework on which this project builds. We gratefully acknowledge the University of Maryland Discovery Fund and TEDCO for their support of the program.
\end{acknowledgments}

\bibliography{references}

@inproceedings{bennett1984,
  author    = {Bennett, Charles H. and Brassard, Gilles},
  title     = {Quantum cryptography: Public key distribution and coin tossing},
  booktitle = {Proceedings of the IEEE International Conference on Computers, Systems and Signal Processing},
  address   = {Bangalore, India},
  pages     = {175--179},
  year      = {1984},
  note      = {Reprinted in Theor. Comput. Sci. \textbf{560}, 7--11 (2014)}
}

@article{cao2021,
  author  = {Cao, Xiao-Yu and Gu, Jie and Lu, Yu-Shuo and Yin, Hua-Lei and Chen, Zeng-Bing},
  title   = {Coherent one-way quantum conference key agreement based on twin field},
  journal = {New Journal of Physics},
  volume  = {23},
  number  = {4},
  pages   = {043002},
  year    = {2021}
}

@article{chen2007,
  author  = {Chen, Kai and Lo, Hoi-Kwong},
  title   = {Multi-partite quantum cryptographic protocols with noisy {GHZ} states},
  journal = {Quantum Information \& Computation},
  volume  = {7},
  number  = {8},
  pages   = {689--715},
  year    = {2007}
}

@article{ekert1991,
  author  = {Ekert, Artur K.},
  title   = {Quantum cryptography based on {Bell}'s theorem},
  journal = {Physical Review Letters},
  volume  = {67},
  number  = {6},
  pages   = {661--663},
  year    = {1991}
}

@article{epping2017,
  author  = {Epping, Michael and Kampermann, Hermann and Macchiavello, Chiara and Bru{\ss}, Dagmar},
  title   = {Multi-partite entanglement can speed up quantum key distribution in networks},
  journal = {New Journal of Physics},
  volume  = {19},
  number  = {9},
  pages   = {093012},
  year    = {2017}
}

@article{grasselli2018,
  author  = {Grasselli, Federico and Kampermann, Hermann and Bru{\ss}, Dagmar},
  title   = {Finite-key effects in multipartite quantum key distribution protocols},
  journal = {New Journal of Physics},
  volume  = {20},
  number  = {11},
  pages   = {113014},
  year    = {2018}
}

@article{grasselli2019,
  author  = {Grasselli, Federico and Kampermann, Hermann and Bru{\ss}, Dagmar},
  title   = {Conference key agreement with single-photon interference},
  journal = {New Journal of Physics},
  volume  = {21},
  number  = {12},
  pages   = {123002},
  year    = {2019}
}

@article{grasselli2022,
  author  = {Grasselli, Federico and Murta, Gl{\'a}ucia and de Jong, Jarn and Hahn, Frederik and Bru{\ss}, Dagmar and Kampermann, Hermann and Pappa, Anna},
  title   = {Secure anonymous conferencing in quantum networks},
  journal = {PRX Quantum},
  volume  = {3},
  number  = {4},
  pages   = {040306},
  year    = {2022}
}

@article{hahn2020,
  author  = {Hahn, Frederik and de Jong, Jarn and Pappa, Anna},
  title   = {Anonymous quantum conference key agreement},
  journal = {PRX Quantum},
  volume  = {1},
  number  = {2},
  pages   = {020325},
  year    = {2020}
}

@article{hillery1999,
  author  = {Hillery, Mark and Bu{\v{z}}ek, Vladim{\'i}r and Berthiaume, Andr{\'e}},
  title   = {Quantum secret sharing},
  journal = {Physical Review A},
  volume  = {59},
  number  = {3},
  pages   = {1829--1834},
  year    = {1999}
}

@article{holz2020,
  author  = {Holz, Timo and Kampermann, Hermann and Bru{\ss}, Dagmar},
  title   = {Genuine multipartite {Bell} inequality for device-independent conference key agreement},
  journal = {Physical Review Research},
  volume  = {2},
  number  = {2},
  pages   = {023251},
  year    = {2020}
}

@article{katabarwa2015,
  author  = {Katabarwa, Amara and Geller, Michael R.},
  title   = {Logical error rate in the {Pauli} twirling approximation},
  journal = {Scientific Reports},
  volume  = {5},
  pages   = {14670},
  year    = {2015}
}

@article{li2021,
  author  = {Li, Zhao and Cao, Xiao-Yu and Li, Chen-Long and Weng, Chen-Xun and Gu, Jie and Yin, Hua-Lei and Chen, Zeng-Bing},
  title   = {Finite-key analysis for quantum conference key agreement with asymmetric channels},
  journal = {Quantum Science and Technology},
  volume  = {6},
  number  = {4},
  pages   = {045019},
  year    = {2021}
}

@misc{memmen2023,
  author        = {Memmen, Janka and Eisert, Jens and Walk, Nathan},
  title         = {Advantage of multi-partite entanglement for quantum cryptography over long and short ranged networks},
  year          = {2023},
  eprint        = {2312.13376},
  archivePrefix = {arXiv},
  primaryClass  = {quant-ph}
}

@misc{memmen2026,
  author        = {Memmen, Janka and Kunzelmann, Julia and Walk, Nathan and Eisert, Jens and Walln{\"o}fer, Julius},
  title         = {Strategy optimization for quantum conference key agreement in asymmetric star networks},
  year          = {2026},
  eprint        = {2605.18677},
  archivePrefix = {arXiv},
  primaryClass  = {quant-ph}
}

@article{murta2020,
  author  = {Murta, Gl{\'a}ucia and Grasselli, Federico and Kampermann, Hermann and Bru{\ss}, Dagmar},
  title   = {Quantum conference key agreement: A review},
  journal = {Advanced Quantum Technologies},
  volume  = {3},
  number  = {11},
  pages   = {2000025},
  year    = {2020}
}

@article{pickston2023,
  author  = {Pickston, Alexander and Ho, Joseph and Ulibarrena, Andr{\'e}s and Grasselli, Federico and Proietti, Massimiliano and Morrison, Christopher L. and Barrow, Peter and Graffitti, Francesco and Fedrizzi, Alessandro},
  title   = {Conference key agreement in a quantum network},
  journal = {npj Quantum Information},
  volume  = {9},
  pages   = {82},
  year    = {2023}
}

@article{proietti2021,
  author  = {Proietti, Massimiliano and Ho, Joseph and Grasselli, Federico and Barrow, Peter and Malik, Mehul and Fedrizzi, Alessandro},
  title   = {Experimental quantum conference key agreement},
  journal = {Science Advances},
  volume  = {7},
  number  = {23},
  pages   = {eabe0395},
  year    = {2021}
}

@article{ribeiro2018,
  author  = {Ribeiro, J{\'e}r{\'e}my and Murta, Gl{\'a}ucia and Wehner, Stephanie},
  title   = {Fully device-independent conference key agreement},
  journal = {Physical Review A},
  volume  = {97},
  number  = {2},
  pages   = {022307},
  year    = {2018}
}

@article{tomita2014,
  author  = {Tomita, Yu and Svore, Krysta M.},
  title   = {Low-distance surface codes under realistic quantum noise},
  journal = {Physical Review A},
  volume  = {90},
  number  = {6},
  pages   = {062320},
  year    = {2014}
}

@article{webb2024,
  author  = {Webb, Joseph W. and Ho, Joseph and Grasselli, Federico and Murta, Gl{\'a}ucia and Pickston, Alexander and Ulibarrena, Andr{\'e}s and Fedrizzi, Alessandro},
  title   = {Experimental anonymous quantum conferencing},
  journal = {Optica},
  volume  = {11},
  number  = {6},
  pages   = {872--875},
  year    = {2024}
}

@article{zhao2020,
  author  = {Zhao, Shuai and Zeng, Pei and Cao, Wen-Fei and Xu, Xin-Yu and Zhen, Yi-Zheng and Ma, Xiongfeng and Li, Li and Liu, Nai-Le and Chen, Kai},
  title   = {Phase-matching quantum cryptographic conferencing},
  journal = {Physical Review Applied},
  volume  = {14},
  number  = {2},
  pages   = {024010},
  year    = {2020}
}

@article{kam2024,
  author  = {Kam, John F. and Kang, Haiyue and Hill, Charles D. and Mooney, Gary J. and Hollenberg, Lloyd C. L.},
  title   = {Characterization of entanglement on superconducting quantum computers of up to 414 qubits},
  journal = {Physical Review Research},
  volume  = {6},
  number  = {3},
  pages   = {033155},
  year    = {2024}
}

\end{document}